\documentclass[preprint,proceedings]{rmaa}
\usepackage{paralist}

\usepackage{psfrag,color}

\SetYear{2003}
\SetConfTitle{The Eighth Texas-Mexico Conference on Astrophysics }

\title{FORMATION AND EVOLUTION OF SELF-INTERACTING DARK MATTER HALOS} 

\author{
  Kyungjin Ahn\altaffilmark{1} 
  and Paul R. Shapiro\altaffilmark{1}}

\altaffiltext{1}{Department of Astronomy, University of Texas at
Austin, Austin, TX 78712}

\shortauthor{AHN \& SHAPIRO}
\shorttitle{Self-Interacting Dark Matter Halos}

\fulladdresses{
\item Kyungjin Ahn and Paul R. Shapiro: The University of Texas at Austin,
  Department of Astronomy, 1 University Station C1400, Austin, TX 78712-0259,
  USA
}

\listofauthors{Kyungjin Ahn \& Paul Shapiro}
\indexauthor{Ahn, K.}
\indexauthor{Shapiro, P. R.}

\abstract{
We study the formation and evolution of self-interacting dark matter
(SIDM) halos. We find analytical, fully
cosmological similarity solutions for their dynamics, which take proper 
account of the collisional interaction of SIDM particles, based on
a fluid approximation derived from the Boltzmann equation.
These similarity solutions are relevant to galactic and cluster halo 
formation in the cold dark matter (CDM) model.
Different solutions arise for different values of the dimensionless
collisionality 
parameter, \( Q\equiv\sigma \rho_{b}r_{s} \), where \( \sigma  \) is the
SIDM particle scattering cross section per unit mass, \( \rho _{b} \)
is the cosmic mean density, and \( r_{s} \) is the shock radius.
For all solutions, a flat-density, isothermal core is present
which grows in size as a fixed fraction of \( r_{s} \),
pumped by cosmological infall.
Accordingly, core collapse must in general be delayed until infall
becomes negligible, contrary to previous analyses based on isolated halos,
which predict core collapse in a Hubble time. 
Our solutions agree with N-body simulations, which match observed
galactic rotation curves if \(\sigma \sim [0.56 - 5.6]\, cm^{2} g^{-1}
\), implying \( Q \sim [6.2 \times 10^{-7} - 3.6 \times
10^{-5}]\). Similar profiles also arise for \( Q \sim [1.37\times
10^{-2} - 1.7 \times 10^{-1}] \), or \(\sigma \sim [1.2\times 10^{4}
- 2.7 \times 10^{4}] \, cm^{2} g^{-1} \), a regime not previously
simulated. 
}
\addkeyword{cosmology:dark matter}
\addkeyword{cosmology: large-scale structure of universe}
\addkeyword{galaxies: kinematics and dynamics}

\begin{document}
% Typeset article header
\maketitle

\section{Introduction}
\label{sec:intro}

The self-interacting dark matter (SIDM) model was proposed by Spergel
\& Steinhardt (2000) to resolve the
discrepancy between the inner density profile of cold dark matter (CDM)
halos found by N-body simulations and observations of galactic
rotation curves. N-body simulations
have a density cusp (\( \rho \propto r^{\beta } \)
where \( \beta  \) ranges from 
-1 (Navarro, Frenk \& White 1997; hereafter ``NFW'')
to -1.5 (Moore et al. 1999; hereafter ``Moore profile'')),
while observations of dark-matter dominated dwarf and LSB galaxies
indicate flat-density (soft) cores.
SIDM particles undergo nongravitational, microscopic interaction which
is strong enough to produce a soft core by heat conduction.
Cosmological N-body simulations which incorporate a finite scattering
cross-section \( \sigma \) for the SIDM particles 
show that this scheme successfully produces
soft cores in halos (e.g. Dav\'{e} et al. 2001; Yoshida et al. 2000b).

We study the formation and evolution of SIDM halos analytically with a
proper treatment of cosmological infall
(see also Ahn \& Shapiro 2002). Previous analytical studies
(Balberg, Shapiro \& Inagaki 2002) and N-body simulations of
\(isolated\) halos (Burkert 2000; Kochanek \& White 2000) neglected
cosmological infall which might delay the core collapse.
We find that, for a realistic range of infall rates, the collapse of
SIDM cores can be completely inhibited.
Our results agree with N-body simulations for nonisolated halos, but
also find that a range of higher scattering cross sections than
previously simulated can also produce soft cores.

\section{Self-Similar Evolution of SIDM Halos}
\label{sec:main1}

We use a fluid approximation derived from the Boltzmann equation to
describe the dynamics of SIDM halos. A spherically symmetric,
arbitrarily collisional system
with isotropic random motions can be described by fluid equations
for an ideal gas with \(\gamma=5/3\) (Bettwieser 1983).

For a collisionless system, collapse is ``adiabatic'' (i.e. no conduction),
and if the initial overdensity has a scale-free power-law
form,
\begin{equation}
\frac{\delta M}{M}\propto M^{-\varepsilon },
\end{equation}
where \(\varepsilon\) is a positive constant,
the object which grows by adiabatic infall is self-similar
(Fillmore \& Goldreich 1984). Heat conduction generally
breaks self-similarity by introducing an additional length (time)
scale, but when \( r_{s} \propto t^2 \), or \(\varepsilon=1/6 \),
self-similarity is preserved even in the presence of heat conduction
by SIDM particles with constant \(\sigma\).

\section{Self-Similar CDM Halos (\(\varepsilon=1/6\); No Conduction)}
\label{sec:main2}
The condition which is required to preserve self-similarity of the
dynamics of SIDM halos deserves separate attention. We find that
the adiabatic solution with \(\varepsilon=1/6\) is well fit, for
example, by an NFW profile with concentration parameter \( 3 \leq
c_{\mathrm{NFW}} \leq 4 \). The inner density slope is \(-1.27\), which is
between \(-1\) (NFW) and \(-1.5\) (Moore profile) (see Fig. 1).

The theory of structure formation from density peaks in the Gaussian
random noise distribution of initial density fluctuations gives
an interesting clue to this correspondence.
According to Hoffman \& Shaham (1985), 
for a power-law power spectrum of initial fluctuations,
\( P(k)\propto k^{n}, \) the initial density
profile of density peaks can be approximated 
as \( \frac{\delta \rho }{\rho }\propto r^{\kappa }, \)
where \( \kappa=3\varepsilon=n+3 \). The value \( \varepsilon=1/6 \)
corresponds to \( \kappa=1/2, \) or \( n=-2.5 \): \( n=-2.5 \) 
roughly corresponds to galaxy-mass
structures in \(\Lambda\)CDM.

\begin{figure}[!t]
  \begin{center}
  \includegraphics[width=5cm]{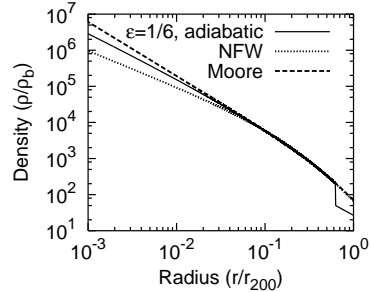}
  \end{center}
  \caption{Halo density profile for adiabatic solution with
  \(\varepsilon=1/6\) for standard CDM halos, compared to NFW and Moore
  profiles. The adiabatic solution, whose inner slope is -1.27, is a
  good fit to density profiles of CDM halos from N-body simulations.
  }
  \label{fig:fig1}
\end{figure}

\section{Self-Similar SIDM Halos (\(\varepsilon=1/6\); With Conduction)}
\label{sec:main3}
We find that different similarity solutions of SIDM halos arise for
different values of the collisionality parameter, 
\( Q\equiv \sigma \rho _{b}r_{s} \).
High \( Q \) means high collision rate. We also find that
there are two different regimes: the low-\( Q \) regime
(\(Q<Q_{th}=7.35\times 10^{-4}\)) and the high-\( Q \)
regime (\(Q>Q_{th}\)).

\begin{figure*}[!t]
  \begin{center}
  \includegraphics[height=6cm]{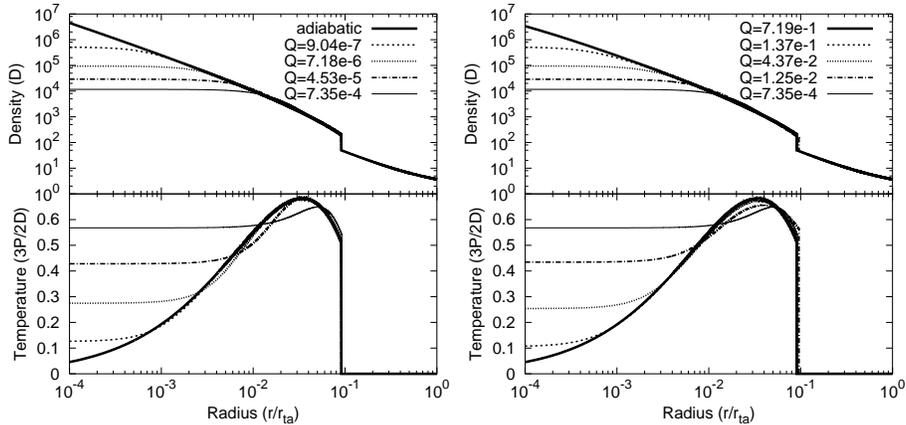}
  \end{center}
  \caption{Similarity solution dimensionless profiles for
  low-\(Q\) (left panels) and high-\(Q\) (right panels) regimes.
  Even though values of \(Q\) are very
  different, profiles of high-\(Q\) solutions are indistinguishable
  from those of low-\(Q\) solutions.}
  \label{fig:fig2}
\end{figure*}

In the low-\( Q \) regime, the collision mean free path in the core
is larger than the size of the halo. In this regime,
the flattening of the core density
profile increases as \( Q \) increases (see Fig. 2). 

In the high-\( Q \) regime, the mean free path in the core region
is smaller than the size of the halo. The flattening of the core density
profile decreases as \( Q \) increases, because the mean free path
decreases (diffusion limit; see Fig. 2). The limiting case of
\( Q=\infty  \) corresponds to the solution
with no conduction, which agrees with N-body simulations with infinite
cross-section (Yoshida et al. 2000a; Moore et al. 2000). 
In this case,
the nonadiabatic solution approaches the adiabatic solution.

Our similarity solutions naturally explain how cosmological infall
affects SIDM halos dynamically. New matter which falls into the halo
from outside gets shock-heated (i.e. virialized) to such an extent
that the temperature is always greater in the halo than in the core.
Accordingly, in our similarity solutions, core collapse is completely
prohibited. Therefore, core collapse will be inhibited in general until the
infall rate becomes negligible at late times in the 
evolution of individual halos, as reported for N-body simulations by
Wechsler et al. (2002).

\section{The Allowed Range of Scattering Cross-Section for an SIDM Universe}
\label{sec:main4}

\begin{figure}[!b]
  \begin{center}
  \includegraphics[height=5cm]{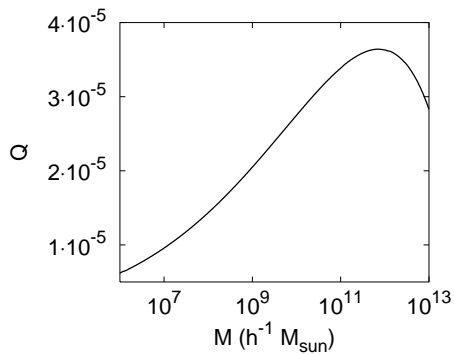}
  \end{center}
  \caption{The collisionality parameter \(Q\) vs. mass of halos at
  their typical formation epoch for \(\sigma=5.6cm^{2} g^{-1}\) and
  \(h=0.7\) in \(\Lambda\)CDM universe. 
  }
  \label{fig:fig3}
\end{figure}

We find that a relatively narrow range of Q values,  
\begin{equation}
\label{eq:qs}
Q\simeq[0.62 - 3.6]\times10^{-5}\left( \frac{h}{0.70}\right) \left( \frac{\sigma }{5.6cm^{2}g^{-1}}\right),
\end{equation}
characterizes the entire range of halo masses from \(10^{6} h^{-1}
M_{\sun}\) to \(10^{13} h^{-1} M_{\sun}\), in the currently-favored
\(\Lambda\)CDM universe (see Fig. 3).

The range \( \sigma =[0.56\, -\, 5.6]\,cm^{2}g^{-1} \) is the preferred range
of the scattering cross section found by cosmological N-body simulations
for the \(\Lambda\)CDM universe
to match observed galactic rotation curves (e.g. {Dav\'{e} et al. 2001}).
Equation~(\ref{eq:qs}) with \( h=0.7 \) then yields 
\( Q=[6.2\times10^{-7}\, -\, 3.6\times 10^{-5}] \),
which is in the low-\( Q \) regime. 

As shown in \S~\ref{sec:main3}, there also exist high-\( Q \)
solutions which yield profiles which are quite similar
to the low-\( Q \) solutions which produce observationally acceptable
soft cores. We find that \( Q=[1.37\times 10^{-2}\, -\, 1.7\times 10^{-1}] \) 
in the high-\(Q\) regime produces profiles similar to those in the low-\(Q\)
regime for
\( Q=[6.2\times10^{-7}\, -\, 3.6\times 10^{-5}] \).
From the relationship between 
\( \sigma  \) and \( Q \) (equation~\ref{eq:qs}), we predict that
\( \sigma=[1.2\times 10^{4}-2.7\times 10^{4}]\,cm^{2}g^{-1} \) can also
produce acceptable soft cores, therefore. Cosmological N-body simulations
which incorporate SIDM particles with \( \sigma \simeq [5\times
10^{3}-5\times 10^{4}]\,cm^{2}g^{-1} \) would, therefore, be of interest.

\acknowledgements

This work was partially supported by NASA ATP Grant NAG5-10825 and Texas
Advanced Research Program Grant 3658-0624-1999.

\end{document}